\documentclass[%
 aip,
 jmp,%
 amsmath,amssymb,
reprint%
]{revtex4-1}

\usepackage{graphicx}
\usepackage{dcolumn}
\usepackage{bm}
\usepackage{subfigure} 

\begin{document}


\title[Line tension induced scenario]{Line-tension-induced scenario of heterogeneous
nucleation on a spherical substrate and in a spherical cavity
}

\author{Masao Iwamatsu}
\email{iwamatsu@ph.ns.tcu.ac.jp}
\affiliation{ 
Department of Physics, Faculty of Liberal Arts and Sciences, Tokyo City University, Setagaya-ku, Tokyo 158-8557, JAPAN
}%

\date{\today}

\begin{abstract}
Line-tension-induced {scenario of heterogeneous nucleation} is studied for a lens-shaped nucleus with a finite contact angle nucleated on a spherical substrate and on the bottom of the wall of a spherical cavity.  The effect of line tension on the free energy of a critical nucleus can be separated from the usual volume term.  By comparing the free energy of a lens-shaped critical nucleus of a finite contact angle with that of a spherical nucleus,  we find that a spherical nucleus may have a lower free energy than a lens-shaped nucleus when the line tension is positive and large, which is similar to the drying transition predicted by Widom [B. Widom, J. Phys. Chem. {\bf 99} 2803 (1995)].  Then, the homogeneous nucleation rather than the heterogeneous nucleation will be favorable.  Similarly, the free energy of a lens-shaped nucleus becomes negative when the line tension is negative and large.  Then, the barrier-less nucleation with no thermal activation called athermal nucleation will be realized. 
\end{abstract}

\pacs{64.60.Q-}
\keywords{Heterogeneous nucleation, Spherical Substrate, Spherical Cavity}
\maketitle

\section{Introduction}
Line tension~\cite{Bonn2009} occurs in the presence of a three phase contact line, which separates three phases such as liquid, solid, and vapor.  Line tension plays a fundamental role in, for example, the stability of liquid droplets adsorbed onto a solid substrate. However, it has been debated whether line tension plays a role in wetting because the magnitude of line tension is quite low~\cite{Gaydos1987,Vera-Graziano1995,Drelich1996,Pompe2000,Wang2001,Checco2003, Bonn2009}. Line tension should play role in heterogeneous nucleation on a substrate as well because three-phase contact exists in such a case~\cite{Kelton2010}. However, the problem of line tension in heterogeneous nucleation has been largely ignored, with the exception of the pioneering theoretical work by Navascu\'es and Tarazona~\cite{Navascues1981} and the recent detailed study by Singha et al.~\cite{Singha2015}.

The heterogeneous nucleation of a lens-shaped nucleus on an ideally spherical substrate was theoretically formulated half a century ago (in 1958) by Fletcher~\cite{Fletcher1958} after the work by Turnbull~\cite{Turnbull1950} on a flat substrate. Since then, this theory has been applied to study, for example, heterogeneous nucleation in the atmosphere~\cite{Lazaridis1991,Gorbunov1997,Hellmuth2013}.  Further theoretical analysis has not been conducted until recently, because of the mathematical complexity of the problem~\cite{Xu2005,Qian2007,Qian2009,Qian2012}. The same problem has also been studied from the standpoint of wetting~\cite{Hage1984}.  However, thus far, few studies have been conducted on the effect of line-tension on a spherical substrate~\cite{Scheludko1985,Hienola2007}, while many studies have been conducted on the effect of line tension on a flat substrate~\cite{Navascues1981,Widom1995,Greer2010,Singha2015}.  There have also been several studies on heterogeneous nucleation in a confined volume~\cite{Cooper2007,Liu2008} and within a cavity~\cite{Maksimov2010,Maksimov2013}.  A similar problem of a macroscopic droplet on convex and concave spherical surfaces has also been studied~\cite{Extrand2012}.  

Recently, line-tension-dominated nucleation has been studied through computer simulation~\cite{Auer2003,Cacciuto2004}.  Furthermore, the importance of the line tension was recently pointed out by Greer~\cite{Greer2010}. 
In fact, nucleation is often heterogeneous and is assisted by the presence of a substrate or wall and impurities. Therefore, understanding the effect of line tension on nucleation is crucial to clarify the whole process of heterogeneous nucleation. However, since a theoretical study of line tension has been hindered by the complex mathematics and geometry,  most recent studies on heterogeneous nucleation have relied on computer simulations~\cite{Auer2003,Cacciuto2004,Sear2008,Liu2011,Sandomirski2014} or ad hoc assumptions~\cite{Ghosh2013}, the predictive power of which for designing new material is limited compared to mathematically rigorous formulations. 

In the present study, we extend our existing knowledge on line tension on convex and concave spherical surfaces~\cite{Iwamatsu2015} and consider the free energy of a lens-shaped critical nucleus nucleated on a spherical substrate and on a wall of a spherical cavity within the framework of the classical nucleation theory (CNT).  We consider the critical nucleus as a continuum of uniform density having a sharp interface with a part of a spherical substrate.  Consequently, the contact line and angle can be defined without ambiguity.  The nucleus-substrate interaction is represented by the appropriate surface tension (energy) confined to the contact area so that the classical concept of surface tension and interfacial energy can be applied. 

We find that a spherical nucleus of homogeneous nucleation may have a lower free energy than a lens-shaped nucleus of heterogeneous nucleation when the line tension is positive and large, which is similar to the drying transition~\cite{Widom1995}.  We also find that the free energy of a lens-shaped nucleus becomes negative when the line tension is negative and large, which will leads to the barrier-less athermal nucleation~\cite{Kelton2010,Quested2005} with no thermal activation process.

\section{\label{sec:sec2}Line-tension effect on the free energy of nucleus}

\subsection{Nucleus on a convex spherical substrate}

In this section, we summarize the mathematical results of our previous work~\cite{Iwamatsu2015} and discuss the physics of line tension in detail.  We consider a lens-shaped liquid nucleus nucleated on a spherical substrate from oversaturated vapor.  However, the result is general and can be applied to the nucleation of crystal grains or vapor bubbles as well.  According to the classical idea of wetting and nucleation theory~\cite{Kelton2010,Fletcher1958,Navascues1981,Qian2009}, the Helmholtz free energy of a nucleus (sessile droplet) is given by
\begin{equation}
\Delta F=\sigma_{\rm lv}A_{\rm lv}+\Delta\sigma A_{\rm sl}+\tau L,
\label{eq:LN1}
\end{equation}
and
\begin{equation}
\Delta\sigma = \sigma_{\rm sl}-\sigma_{\rm sv},
\label{eq:LN2}
\end{equation}
where $A_{\rm lv}$ and $A_{\rm sl}$ are the surface areas of the liquid-vapor and liquid-solid (substrate) interfaces, respectively, and $\sigma_{\rm lv}$ and $\sigma_{\rm sl}$ are their respective surface tensions.  Moreover, $\Delta \sigma$ is the free energy gained when the solid-vapor interface with surface tension $\sigma_{\rm sv}$ is replaced by the solid-liquid interface with surface tension $\sigma_{\rm sl}$.   The effect of the line tension $\tau$ is given by the last term, where $L$ denotes the length of the three-phase contact line.

\begin{figure}[htbp]
\begin{center}
\includegraphics[width=0.80\linewidth]{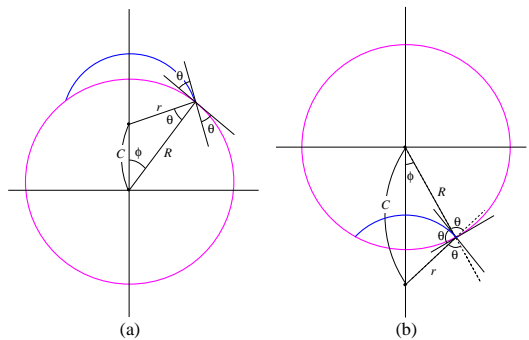}
\caption{
(a) Liquid nucleus on a convex substrate.  The centers of the nucleus with radius $r$ and that of the spherical substrate or cavity with radius $R$ are separated by a distance $C$.  The contact angle is denoted by $\theta$.
The angles $\phi$ and $\theta$ are related by $C\sin\phi = r\sin\theta$, $C\cos\phi = R-r\cos\theta$.  Two radii $R$ and $r$ are related to the distance $C$ through $C^2=R^{2}+r^{2}-2Rr\cos\theta$.  (b)  Liquid nucleus on a concave substrate of a spherical cavity.  The angles $\phi$ and $\theta$ are related by $C\sin\phi = r\sin\theta$, $C\cos\phi = R+r\cos\theta$.  Two radii $R$ and $r$ are related to the distance $C$ through $C^2=R^{2}+r^{2}+2Rr\cos\theta$.   Note that the three-phase contact line passes through the equator when $\phi=90^{\circ}$ and the contact line moves from the upper (lower) hemisphere to lower (upper) hemisphere on the sphere (cavity).  }
\label{fig:LN1}
\end{center}
\end{figure}

The contact angle is determined by minimizing the Helmholtz free energy in Eq.~(\ref{eq:LN1}) under the condition that the nucleus volume is constant at $V$.  By changing the variable  $\theta$ to the distance $C$ between two spheres defined in Fig.~\ref{fig:LN1}, it is possible to minimize the free energy in Eq.~(\ref{eq:LN1}) with respect to the radius $r$.  Detail of this calculation can be found elsewhere~\cite{Iwamatsu2015}.  We finally obtain the generalized Young equation that determines  the contact angle $\theta$ as~\cite{Iwamatsu2015}
\begin{equation}
\Delta\sigma+\sigma_{\rm lv}\cos\theta+\frac{R-r\cos\theta}{Rr\sin\theta}\tau=0,
\label{eq:LN3}
\end{equation}
which is similar to the classical Young equation~\cite{Young1805} on a flat substrate,
\begin{equation}
\Delta\sigma+\sigma_{\rm lv}\cos\theta_{0}=0,
\label{eq:LN4}
\end{equation}
where $\theta_{0}$ is the classical Young's contact angle, while the contact angle $\theta$ is the intrinsic contact angle~\cite{Marmur1998} of the nucleus.  Equation (\ref{eq:LN3}) has also been derived from the general theory of differential geometry by using the geodesic curvature~\cite{Guzzardi2007}.  Even on a spherical curved surface, the contact angle will be determined from the classical Young equation (Eq.~(\ref{eq:LN4})) for flat surfaces~\cite{Fletcher1958,Qian2009} when the line tension can be neglected ($\tau=0$).

Equation (\ref{eq:LN3}) can also be written as~\cite{Iwamatsu2015}
\begin{equation}
\sigma_{\rm lv}\cos\theta=\sigma_{\rm lv}\cos\theta_{0}-\frac{\tau}{R\tan\phi},
\label{eq:LN5}
\end{equation}
using the angle $\phi$ defined in Fig.~\ref{fig:LN1}.  Equation (\ref{eq:LN5}) is known as the generalized Young equation~\cite{Hienola2007}. This formula is different from that proposed by Scheludko~\cite{Scheludko1985}, but is the same as that proposed by Hienola~\cite{Hienola2007}.  Equation (\ref{eq:LN5}) can be derived also from the local approach~\cite{Roura2005} rather than the global approach of minimizing the Helmholtz free energy~\cite{Iwamatsu2015,Guzzardi2007}.

\begin{figure}[htbp]
\begin{center}
\subfigure[Tension $\sigma_{\tau}$ resulting from the line tension $\tau$.]
{
\includegraphics[width=0.45\linewidth]{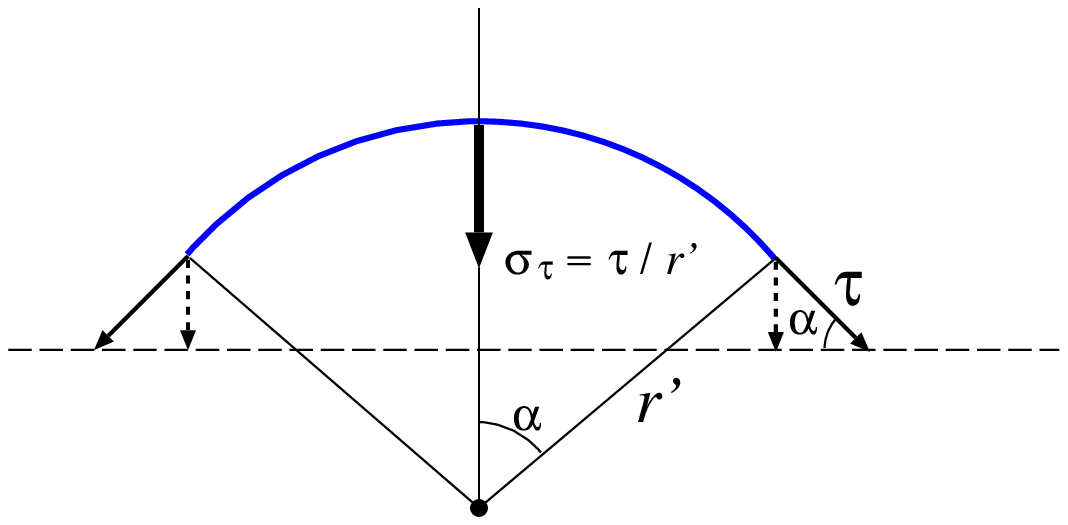}
\label{fig:LN2a}
}
\subfigure[Mechanical force balance of the surface tensions and line tension.]
{
\includegraphics[width=0.45\linewidth]{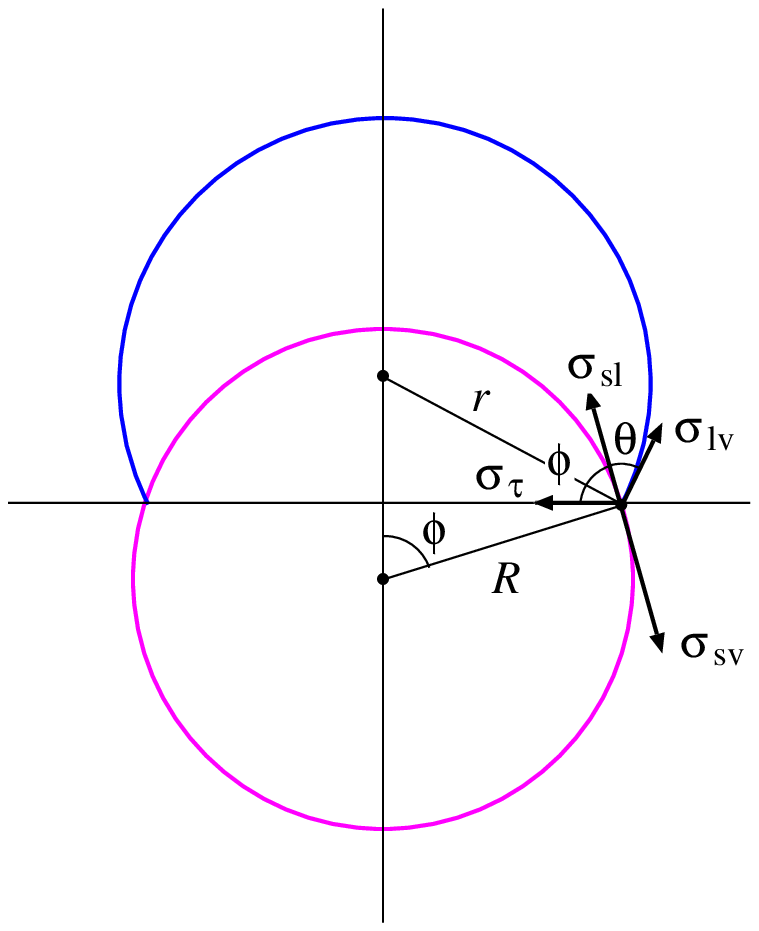}
\label{fig:LN2b}
}
\end{center}
\caption{
(a) Tension $\sigma_{\tau}$ from the line tension $\tau$ exerted on a portion of the contact line of arc length $2\alpha r^{'}$.  The tension $\sigma_{\tau}$ on unit length  is given by Eq.~(\ref{eq:LN12}).  (b) The mechanical force balance among the three surface tensions $\sigma_{\rm lv}$, $\sigma_{\rm sl}$, and $\sigma_{\rm sv}$ and the tension $\sigma_{\tau}$ from the line tension $\tau$ for the nucleus on a convex spherical substrate.
 } 
\label{fig:LN2}
\end{figure}

In fact, Eq.~(\ref{eq:LN5}) can be derived from the mechanical force balance of the surface tension as pointed out by Hienola {\it et al.}~\cite{Hienola2007}.  To this end, we first note that the line tension contributes to the force balance as (Fig.~\ref{fig:LN2a})
\begin{equation}
\sigma_{\tau}=\lim_{\alpha\rightarrow 0}\frac{2\tau\sin\alpha}{2\alpha r^{'}}=\frac{\tau}{r^{'}}.
\label{eq:LN6}
\end{equation}
The line tension contributes to the tension $\sigma_{\tau}$ at the three-phase contact line only when the contact line has the curvature $r^{'}$.  Then, a simple force balance between the three tensions $\sigma_{\rm lv}$, $\sigma_{\rm sl}$, and $\sigma_{\rm sv}$ and $\sigma_{\tau}$ (Fig.~\ref{fig:LN2b}) leads to
\begin{equation}
\sigma_{\rm sl}-\sigma_{\rm sv}+\sigma_{\rm lv}\cos\theta +\sigma_{\tau}\cos\phi=0,
\label{eq:LN7}
\end{equation}
which will be reduced to Eq.~(\ref{eq:LN5}) since $\sigma_{\tau}=\tau/R\sin\phi$. This equation was originally derived for the critical nucleus of heterogeneous nucleation~\cite{Hienola2007}.  However, it is apparent that Eq.~(\ref{eq:LN5}) is also applicable to, for example, the contact angle of a droplet of non-volatile liquids of any size.

Equation (\ref{eq:LN6}) also suggests that the limit $r^{'}\rightarrow 0$ is unphysical.  The line tension $\tau$ must inevitably be curvature dependent; otherwise, divergence occurs when $r^{'}\rightarrow 0$ or $\phi\rightarrow 0^{\circ}$ and $\phi\rightarrow 180^{\circ}$ in Eq.~(\ref{eq:LN5}).  In fact, in the limits $\phi\rightarrow 0^{\circ}$ and $\phi\rightarrow 180^{\circ}$, the substrate must be covered by a thin liquid layer before the perimeter of the contact line disappears.  This is similar to the limit $\theta_{0}\rightarrow 0^{\circ}$ of the classical Young equation (Eq.~(\ref{eq:LN4})), which does not imply that the nucleus disappears. Rather, it implies that the bare substrate with the surface energy $\sigma_{\rm sv}$ will be covered by a thin wetting layer of free energy $\sigma_{\rm sl}+\sigma_{\rm lv}$.  Therefore, the limits $\phi\rightarrow 0^{\circ}$ and $\phi\rightarrow 180^{\circ}$ needs caution and will be unphysical in Eq.~(\ref{eq:LN5}) and in subsequent discussions.  It must also be noted that the line-tension effect in Eq.~(\ref{eq:LN5}) changes its sign at $\phi=90^{\circ}$.  Therefore, the line tension acts in the opposite direction on the upper and lower hemispheres.  This fact can be easily understood as the advancement of the contact line leads to the increase in the nucleus perimeter on the upper hemisphere, while it leads to the decrease in the perimeter on the lower hemisphere.

\begin{figure}[htbp]
\begin{center}
\includegraphics[width=0.45\linewidth]{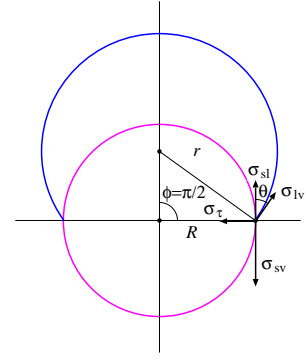}
\caption{
Nucleus for which the three-phase contact line coincides with the equator of the spherical substrate.  In this case, the line tension contributes only to the free energy and not to the mechanical force balance  through $\sigma_{\tau}$.  The contact angle will be pinned at the value given by Eq.~(\ref{eq:LN8}) irrespective of the magnitude of line tension $\tau$.
 }
\label{fig:LN3}
\end{center}
\end{figure}

When the angle $\phi$ is given by $\phi=90^{\circ}$, Eq.~(\ref{eq:LN5}) indicates that the line tension does not contribute to the contact angle $\theta$ and that it remains at the Young's contact angle $\theta_{0}$ irrespective of the magnitude of line tension.  In this case, the contact line is on the equator of the spherical substrate (Fig.~\ref{fig:LN3}), and this is the neutral line because both the advancement and retardation of the contact line leads to the decrease in the nucleus perimeter.  Then, the line tension cannot change the contact line and, therefore, cannot affect the contact angle.  It only affects the free energy through Eq.~(\ref {eq:LN1}).  The characteristic contact angle $\theta_{\rm c}$ that corresponds to $\phi=90^{\circ}$  satisfies  (see Fig.~\ref{fig:LN1}) 
\begin{equation}
\cos\theta_{\rm c}=\frac{R}{r}.
\label{eq:LN8}
\end{equation}
Obviously, $r$ must be larger than $R$; otherwise, the contact line cannot reach the equator.  The contact angle $\theta_{\rm c}<90^{\circ}$ when $\phi=90^{\circ}$.  Therefore, the convex substrate must be hydrophilic so that the contact line crosses the equator.  We will use the popular terminology hydrophobic and hydrophilic throughout, though the liquid is not necessarily water. 

In the limit of infinite substrate radius ($R\rightarrow\infty$), we can recover the modified Young equation~\cite{Navascues1981},
\begin{equation}
\Delta\sigma+\sigma_{\rm lv}\cos\theta+\frac{1}{r\sin\theta}\tau=0,
\label{eq:LN9}
\end{equation}
or
\begin{equation}
\sigma_{\rm lv}\cos\theta=\sigma_{\rm lv}\cos\theta_{0}-\frac{\tau}{r\sin\theta},
\label{eq:LN10}
\end{equation}
on the basis of Eq.~(\ref{eq:LN3}) for a nucleus on a flat substrate.

The critical radius $r$ of the critical nucleus is determined by maximizing the Gibbs free energy of formation~\cite{Navascues1981}
\begin{equation}
\Delta G=\Delta F-\Delta p V,
\label{eq:LN11}
\end{equation}
where $\Delta p$ is the excess vapor pressure of the oversaturated vapor relative to the saturated pressure, $\Delta F$ is the Helmholtz free energy given by Eq.~(\ref{eq:LN1}), and  $V$ is the nucleus volume.  By maximizing Eq.~(\ref{eq:LN11}) with respect to the radius $r$ under the condition of constant contact angle $\theta$, we obtain~\cite{Iwamatsu2015} the well-known Young-Laplace formula 
\begin{equation}
r_{*}=\frac{2\sigma_{\rm lv}}{\Delta p}
\label{eq:LN12}
\end{equation}
for the critical radius of the nucleus. 

By inserting the critical radius $r_{*}$ into Eq.~(\ref{eq:LN11}) and using the generalized Young equation (Eq.~(\ref{eq:LN3})), we obtain the work of formation~\cite{Iwamatsu2015}
\begin{equation}
\Delta G^{*}=\Delta G_{\rm vol}^{*}+\Delta G_{\rm lin}^{*},
\label{eq:LN13}
\end{equation}
which consists of the volume term $\Delta G_{\rm vol}^{*}$ and the line term $\Delta G_{\rm lin}^{*}$. 
The former can be written as
\begin{equation}
\Delta G_{\rm vol}^{*} = f_{\rm cv}\left(\rho,\theta\right)\Delta G_{\rm homog}^{*}.
\label{eq:LN14}
\end{equation}
where
\begin{equation}
\rho = \frac{r_{*}}{R}
\label{eq:LN17}
\end{equation}
is the size factor of the critical nucleus and $\theta$ is the contact angle of the critical nucleus determined from the critical radius $r_{*}$ through Eq.~(\ref{eq:LN3}).  Note that the limit $\rho\rightarrow 0$ corresponds to a flat substrate, and $\rho\rightarrow \infty$ represents a nucleus with a point impurity, which corresponds to homogeneous nucleation with the work of formation given by
\begin{equation}
\Delta G_{\rm homog}^{*}=\frac{2\pi}{3}r_{*}^{3}\Delta p.
\label{eq:LN15}
\end{equation}
The generalized shape factor $f_{\rm cv}\left(\rho,\theta\right)$ for the nucleus on a convex spherical substrate is given by~\cite{Iwamatsu2015}
\begin{eqnarray}
f_{\rm cv}\left(\rho,\theta\right)&=&\frac{1}{4\rho^{3}}
\left(1+2\rho-\sqrt{1+\rho^{2}-2\rho\cos\theta}\right) \nonumber \\
&&\times\left(-1+\rho+\sqrt{1+\rho^{2}-2\rho\cos\theta}\right)^{2},
\label{eq:LN16}
\end{eqnarray}
which reduces to the well-known shape factor originally derived by Fletcher~\cite{Fletcher1958} after tedious manipulation of algebra.  It also reduces to the shape factor~\cite{Turnbull1950,Navascues1981}
\begin{equation}
f_{\rm cv}\left(\rho\rightarrow 0,\theta\right)=\frac{\left(2+\cos\theta\right)\left(1-\cos\theta\right)^2}{4}
\label{eq:LN18}
\end{equation}
for a flat substrate ($\rho\rightarrow 0$).   This shape factor Eq.~(\ref{eq:LN16}) also has a correct limit of homogeneous nucleation $f_{\rm cv}\rightarrow 1$ when $\rho\rightarrow \infty$ and $\theta\rightarrow 180^{\circ}$.

The line contribution $\Delta G_{\rm lin}^{*}$ in Eq.~(\ref{eq:LN13}) for the nucleus on a convex spherical substrate can be written as~\cite{Iwamatsu2015}
\begin{equation}
\Delta G_{\rm lin}^{*}=2\pi r_{*}\tau {\rm g}_{\rm cv}\left(\rho,\theta\right)
\label{eq:LN19}
\end{equation}
using the generalized shape factor for the line contribution~\cite{Iwamatsu2015}
\begin{equation}
{\rm g}_{\rm cv}\left(\rho,\theta\right)
=\frac{-1+\rho\cos\theta+\sqrt{1+\rho^{2}-2\rho\cos\theta}}
{\rho^{2}\sin\theta},
\label{eq:LN20}
\end{equation}
which reduces to
\begin{equation}
{\rm g}_{\rm cv}\left(\rho\rightarrow 0,\theta\right)=\frac{\sin\theta}{2}
\label{eq:LN21}
\end{equation}
for a flat substrate~\cite{Navascues1981}.  It also has a correct limit of homogeneous nucleation when $\rho\rightarrow \infty$ and $\theta\rightarrow 180^{\circ}$.  Equation (\ref{eq:LN19}) can also be written simply as
\begin{equation}
{\rm g}_{\rm cv}\left(\rho, \phi\right)=\frac{1-\cos\phi}{\rho\sin\phi}
\label{eq:LN22}
\end{equation}
as a function of $\phi$ instead of $\theta$.  In contrast to Eq.~(\ref{eq:LN5}), the sign will not change at $\phi=90^{\circ}$ in Eq.~(\ref{eq:LN22}).  Therefore, at $\phi=90^{\circ}$ or when the contact angle is given by Eq. (\ref{eq:LN8}), the contact line coincides with the equator.  Then, the line tension does not change the contact angle, but it changes the free energy or the energy barrier.  However, the limit $\phi\rightarrow 180^{\circ}$ is unphysical.

Note that the line contribution $\Delta G^{*}_{\rm lin}$ or Eq.~(\ref{eq:LN22}) is not one-half of the line contribution $\tau L/2=\pi \tau R\sin\phi$ to the Helmholtz free energy in Eq.~(\ref{eq:LN1}) except for the case of a flat substrate when $\rho\rightarrow 0$.  In fact, the line contribution in $\Delta F$  in Eq. (\ref{eq:LN1}) becomes $\tau L/2=\pi \tau r_{*}\sin\theta$ for a flat  substrate~\cite{Navascues1981}, which is the same as  $\Delta G^{*}_{\rm lin}$ in Eq.~(\ref{eq:LN19}) with ${\rm g}_{\rm cv}$ given by Eq.~(\ref{eq:LN21}).  Therefore, the line contribution $\Delta G^{*}_{\rm lin}$ is not directly proportional to the contact-line length $L$, except on a flat substrate~\cite{Navascues1981}.

\subsection{Nucleus on a concave spherical substrate}

\begin{figure}[htbp]
\begin{center}
\includegraphics[width=0.5\linewidth]{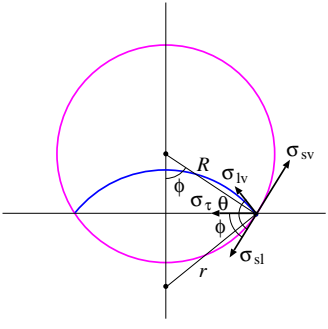}
\caption{
Mechanical force balance among the three surface tension $\sigma_{\rm lv}$, $\sigma_{\rm sl}$, and $\sigma_{\rm sv}$ and the tension $\sigma_{\tau}$ from the line tension $\tau$ for the nucleus on a concave spherical substrate. Note that the contact line is located on the lower hemisphere, whereas it is located on the upper hemisphere in Fig.~\ref{fig:LN2} for a convex substrate.  }
\label{fig:LN4}
\end{center}
\end{figure}

It is also possible to study the contact angle of a nucleus nucleated on a concave spherical substrate (cavity), such as the one shown in Fig.~\ref{fig:LN1}(b)~\cite{Iwamatsu2015}. The contact angle is determined by minimizing the Helmholtz free energy in Eq.~(\ref{eq:LN1}) under the condition of a constant nucleus volume.  The details of this calculation can be found elsewhere~\cite{Iwamatsu2015}.  By using the same procedure as that used in the previous subsection, we arrive at~\cite{Iwamatsu2015} 
\begin{equation}
\Delta\sigma+\sigma_{\rm lv}\cos\theta+\frac{R+r\cos\theta}{Rr\sin\theta}\tau=0,
\label{eq:LN23}
\end{equation}
which can be obtained by changing $\theta\rightarrow 180^{\circ}-\theta$ in Eq.~(\ref{eq:LN3}). Therefore, the action of the line tension on the {\it upper hemisphere of the convex substrate} is the same as that on the {\it lower hemisphere of the concave substrate} (cf. Figs.~\ref{fig:LN2} and \ref{fig:LN4}).  Equation~(\ref{eq:LN23}) can also be written as Eq.~(\ref{eq:LN5}), which can be derived from the mechanical force balance, as shown in Fig.~\ref{fig:LN4}.  Therefore, the role of the line tension changes at $\phi=90^{\circ}$ on a concave substrate as well.

At $\phi=90^{\circ}$, the contact line coincides with the equator of the spherical substrate in this case as well.  The line tension does not contribute to the contact angle $\theta$, which remains at the Young's contact angle $\theta_{0}$ irrespective of the magnitude of line tension. The characteristic contact angle $\theta_{\rm c}$ that corresponds to $\phi=90^{\circ}$  is given by
\begin{equation}
\cos\theta_{\rm c}=-\frac{R}{r}.
\label{eq:LN24}
\end{equation}
on the basis of Eq.~(\ref{eq:LN23}).  Obviously, $r$ must be larger than $R$ and $\theta>90^{\circ}$.  Therefore, the substrate must be hydrophobic so that the contact line crosses the equator from the lower hemisphere to the upper hemisphere.

By maximizing the Gibbs free energy (Eq.~(\ref{eq:LN11})) for a nucleus on a concave substrate, we obtain the Young-Laplace equation (Eq.~(\ref{eq:LN12})) for the critical radius~\cite{Iwamatsu2015}.  Therefore, on a concave spherical substrate, the critical radius of a nucleus is equal to that on a flat substrate, even if the effect of line tension is included, because the effect of the substrate, including that of line tension, is confined to the contact surface.   By inserting the critical radius $r_{*}$ given by Eq.~(\ref{eq:LN12}) into Eq.~(\ref{eq:LN11}) and using the generalized Young equation (Eq.~(\ref{eq:LN23})), we obtain the work of formation in the form of Eq.~(\ref{eq:LN13}), which can be written as Eqs.~(\ref{eq:LN14}) and (\ref{eq:LN19}) by replacing the shape factor for a convex substrate $f_{\rm cv}$ and ${\rm g}_{\rm cv}$ with the corresponding values for a concave substrate $f_{\rm cc}$ and ${\rm g}_{\rm cc}$.

The generalized shape factor $f_{\rm cc}\left(\rho,\theta\right)$ for a nucleus on a concave substrate is now given by~\cite{Iwamatsu2015}
\begin{eqnarray}
f_{\rm cc}\left(\rho,\theta\right)&=&\frac{1}{4\rho^{3}}
\left(-1+2\rho+\sqrt{1+\rho^{2}+2\rho\cos\theta}\right) \nonumber \\
&&\times\left(1+\rho-\sqrt{1+\rho^{2}+2\rho\cos\theta}\right)^{2},
\label{eq:LN25}
\end{eqnarray}
which reduces to the shape factor on a concave surface recently derived by Qian and Ma~\cite{Qian2012}.   It also reduces to the well-known shape factor~\cite{Navascues1981} in Eq.~(\ref{eq:LN18}) for a flat substrate ($\rho\rightarrow 0$).   Note that $f_{\rm cc}\left(\rho\rightarrow \infty,\theta\right)=0$ because the nucleus is confined within an infinitesimally small spherical cavity.

The line contribution $\Delta G_{\rm lin}^{*}$ in Eq.~(\ref{eq:LN13}) can also be written as Eq.~(\ref{eq:LN19}).  The generalized shape factor of the line contribution for a nucleus on a concave substrate is given by~\cite{Iwamatsu2015}
\begin{equation}
{\rm g}_{\rm cc}\left(\rho,\theta\right)
=\frac{-1-\rho\cos\theta+\sqrt{1+\rho^{2}+2\rho\cos\theta}}
{\rho^{2}\sin\theta},
\label{eq:LN26}
\end{equation}
which reduces to
\begin{equation}
{\rm g}_{\rm cc}\left(\rho\rightarrow 0,\theta\right)=\frac{\sin\theta}{2}
\label{eq:LN27}
\end{equation}
for a flat substrate~\cite{Navascues1981}, and ${\rm g}_{\rm cc}\rightarrow 0$ when $\rho\rightarrow \infty$ or $\theta\rightarrow 180^{\circ}$ as the three-phase contact line vanishes.   

Equation (\ref{eq:LN26}) can be written simply as Eq.~(\ref{eq:LN22}) as a function of $\phi$ instead of $\theta$.  It can be observed that the generalized shape factor ${\rm g}_{\rm cc}$ on a concave substrate  given by Eq.~(\ref{eq:LN25}) can be obtained by replacing with $\theta$ with $\rightarrow 180^{\circ}-\theta$ in Eq.~(\ref{eq:LN19}) for ${\rm g}_{\rm cv}$ on the convex substrate.  Therefore, the hydrophobicity and hydrophilicity interchange their roles in the line-tension contribution to the free energy between convex and concave substrates.  Furthermore, the upper and lower hemisphere interchanges their role.  Therefore, the effect of line tension on the nucleus on the upper hemisphere of a spherical convex substrate is the same as that on the nucleus on the lower hemisphere of a spherical concave substrate of a cavity.

\section{\label{sec:sec3}Results and discussion }

\subsection{Nucleus on a convex spherical substrate}

The work of formation of a lens-shaped critical nucleus heterogeneously nucleated on a convex spherical substrate is expressed as
\begin{equation}
\Delta G^{*}_{\rm cv, hetero}=\Delta G^{*}_{\rm homog}h_{\rm cv}\left(\rho,\theta\right),
\label{eq:LN28}
\end{equation}
from Eq.~(\ref{eq:LN13}), (\ref{eq:LN14}) and (\ref{eq:LN19}), where
\begin{equation}
h_{\rm cv}\left(\rho,\theta\right) = f_{\rm cv}\left(\rho,\theta\right)+\frac{3}{2}\bar{\tau}{\rm g}_{\rm cv}\left(\rho,\theta\right)
\label{eq:LN29}
\end{equation}
is the scaled work of formation (nucleation barrier), and
\begin{equation}
\bar{\tau}=\frac{\tau}{\sigma_{\rm lv}r_{*}}
\label{eq:LN30}
\end{equation}
is the scaled line tension. Here, the radius $r_{*}$ is fixed from the Young-Laplace equation (Eq.~(\ref{eq:LN12})).  It is well known that the magnitude of line tension is as low~\cite{Pompe2000,Wang2001,Checco2003, Bonn2009} as $\left|\tau\right|\simeq10^{-13}-10^{-9}$ N.  Then, the magnitude of scaled line tension is of the other
\begin{equation}
\left|\bar{\tau}\right|\simeq 10^{-4}-10^{0}
\label{eq:LN31}
\end{equation}
for a nucleus with $\sigma_{\rm lv}\simeq 73$ mN/m (water) and $r_{*}=$10 nm.  
Since the magnitude of the shape factors are~\cite{Iwamatsu2015} $f_{\rm cv}\simeq 1$ and $g_{\rm cv}\simeq 1$, the estimation in Eq.~(\ref{eq:LN31}) is reasonable as the volume term $f_{\rm cv}$ is still dominant in Eq.~(\ref{eq:LN29}).  If we use the much larger estimate $\tau\simeq 10^{-5}$ N from the old experiments for millimeter-sized droplet~\cite{Gaydos1987,Vera-Graziano1995,Drelich1996}, the line contribution in Eq.~(\ref{eq:LN29}) would dominate, which is difficult to imagine.

\begin{figure}[htbp]
\begin{center}
\subfigure[Energy barrier $h_{\rm cv}$ for positive line tension ($\bar{\tau}=0.2$)]
{
\includegraphics[width=0.7\linewidth]{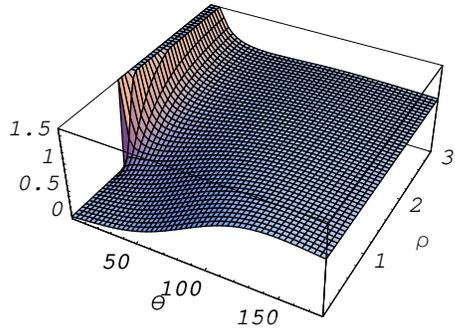}
\label{fig:LN5a}
}
\subfigure[Energy barrier when the line tension is negative ($\bar{\tau}=-0.3$)]
{
\includegraphics[width=0.7\linewidth]{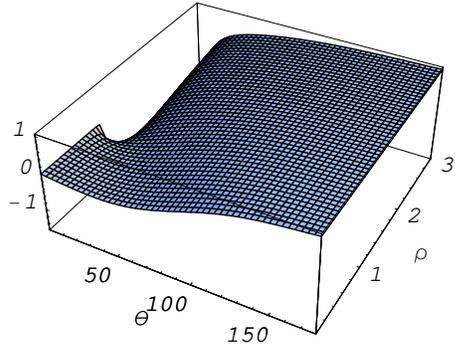}
\label{fig:LN5b}
}
\end{center}
\caption{
Scaled free energy (energy barrier) $h_{\rm cv}\left(\rho,\theta\right)$ of the critical nucleus on a convex spherical substrate as a function of the contact angle $\theta$ and size parameter $\rho$ when the scaled line tension is (a) positive ($\bar{\tau}=0.2$) and (b) negative ($\bar{\tau}=-0.3$). 
 } 
\label{fig:LN5}
\end{figure}

Figure \ref{fig:LN5} shows the scaled energy barrier  $h_{\rm cv}\left(\rho,\theta\right)$ as a functions of the contact angle $\theta$ and size parameter $\rho$ for a heterogeneous nucleus on a convex spherical substrate when $\bar{\tau}=2.0>0$ and $\bar{\tau}=-3.0<0$.  Note that the nucleation rate (probability) $J$ is given by 
\begin{equation}
J \propto \exp\left(-\Delta G^{*}\right/k_{B}T)
\label{eq:LN32}
\end{equation}
and a positive Gibbs free energy $\Delta G^{*}$ acts as the energy barrier of thermal activation with thermal energy $k_{\rm B}T$.  If $\Delta G^{*}$ is negative, the thermal activation is unnecessary and the athermal nucleation~\cite{Quested2005} is realized.

It can be seen from Fig. \ref{fig:LN5a} that the energy barrier can exceed $h=1$, which corresponds to homogeneous nucleation from Eq.~(\ref{eq:LN28}).  Therefore, when $\bar{\tau}>0$, and $\theta$ and $\rho$ are larger, the scaled free energy barrier of the heterogeneous nucleation can exceed that of the homogeneous nucleation. Then, the nucleation rate $J$ in Eq.~(\ref{eq:LN32}) of the homogeneous nucleation will be higher than that of the heterogeneous nucleation. And, the homogeneous nucleation with the spherical critical nucleus rather than the heterogeneous nucleation with the lens-shaped critical nucleus will be favorable, though they can coexist~\cite{Iwamatsu2007}.  This heterogeneous-dominant to homogeneous-dominant change is similar to the drying transition predicted by Widom~\cite{Widom1995}.  On the other hand, when $\bar{\tau}<0$, and $\theta$ and $\rho$ are small in Fig. \ref{fig:LN5b}, the scaled energy barrier becomes negative.  Then, the nucleation becomes deterministic called athermal nucleation because the nucleation does not involve thermal activation to cross the energy barrier~\cite{Quested2005}.

The heterogeneous-dominant to homogeneous-dominant change of a critical nucleus occurs when $h_{\rm cv}\left(\rho,\theta\right)=1$ (Fig.~\ref{fig:LN5a}), which gives the upper bound $\bar{\tau}_{\rm u,cv}$ of the scaled line tension
\begin{equation}
\bar{\tau}_{\rm u,cv}\left(\rho,\theta\right)=\frac{2\left(1- f_{\rm cv}\left(\rho,\theta\right)\right)}{3{\rm g}_{\rm cv}\left(\rho,\theta\right)}.
\label{eq:LN33}
\end{equation}
Similarly, the transition from activated nucleation to non-activated athermal nucleation with negative energy barrier occurs when  $h_{\rm cv}\left(\rho,\theta\right)=0$ (Fig.~\ref{fig:LN5b}), which gives the lower bound $\bar{\tau}_{\rm l,cv}$ of the scaled line tension
\begin{equation}
\bar{\tau}_{\rm l,cv}\left(\rho,\theta\right)=\frac{-2f_{\rm cv}\left(\rho,\theta\right)}{3{\rm g}_{\rm cv}\left(\rho,\theta\right)}.
\label{eq:LN34}
\end{equation}
Since the contact angle $\theta$ will be determined from Eq.~(\ref{eq:LN3}), the scaled line tension $\bar{\tau}$ is given by
\begin{equation}
\bar{\tau}_{\rm cv}\left(\rho,\theta\right)=\frac{\left(\cos\theta_{0}-\cos\theta\right)\sin\theta}{1-\rho\cos\theta},
\label{eq:LN35}
\end{equation}
where $\theta_{0}$ defined by Eq.~(\ref{eq:LN4}) is the Young's contact angle characterizing the substrate with $\sigma_{\rm sl}$ and $\sigma_{\rm sv}$.  

\begin{figure}[htbp]
\begin{center}
\subfigure[Scaled line tension $\bar{\tau}_{\rm cv}$ as a function of the contact angle $\theta$.]
{
\includegraphics[width=0.7\linewidth]{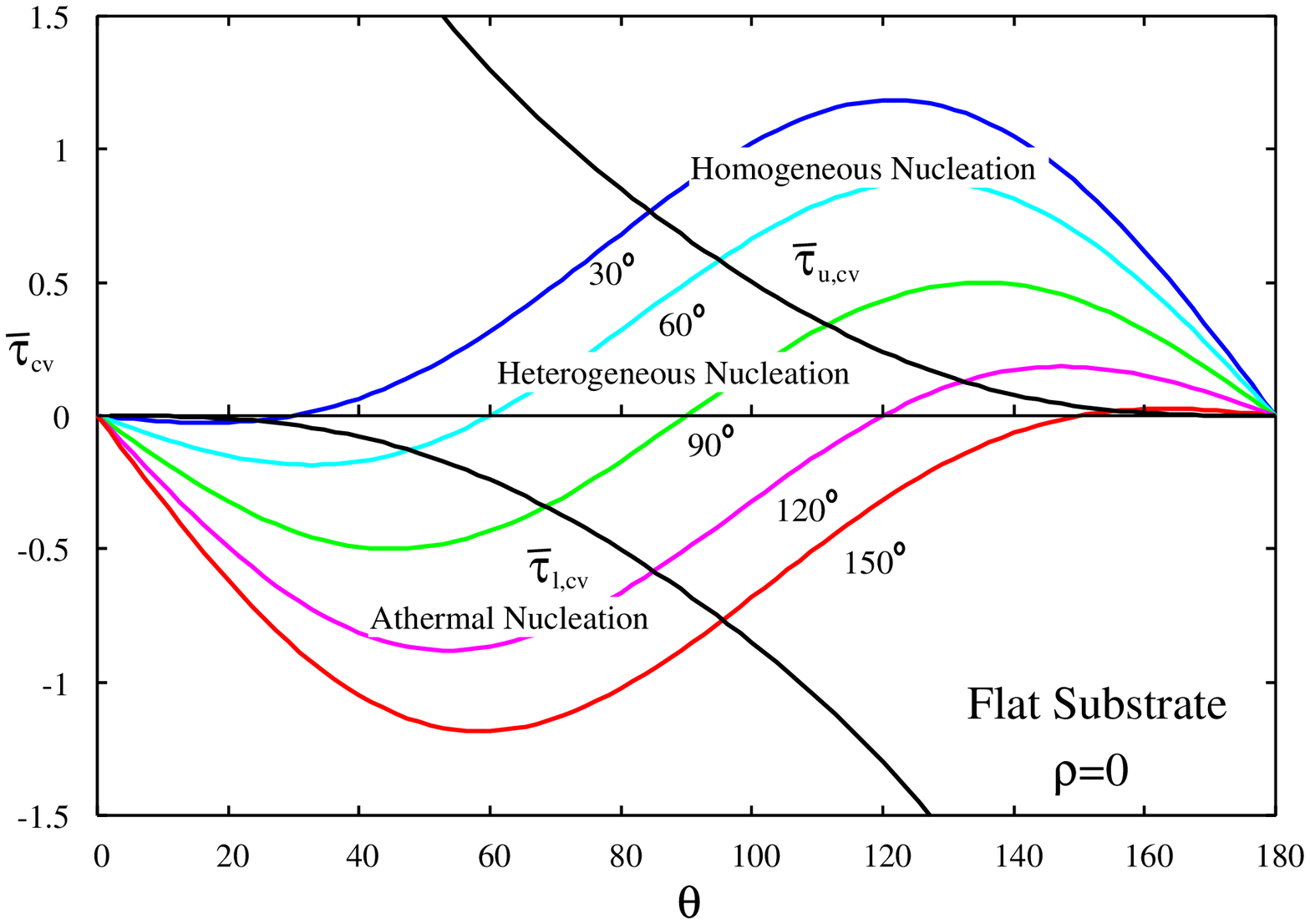}
\label{fig:LN6a}
}
\subfigure[Scaled energy barrier $h_{\rm cv}$ as a function of the contact angle $\theta$]
{
\includegraphics[width=0.7\linewidth]{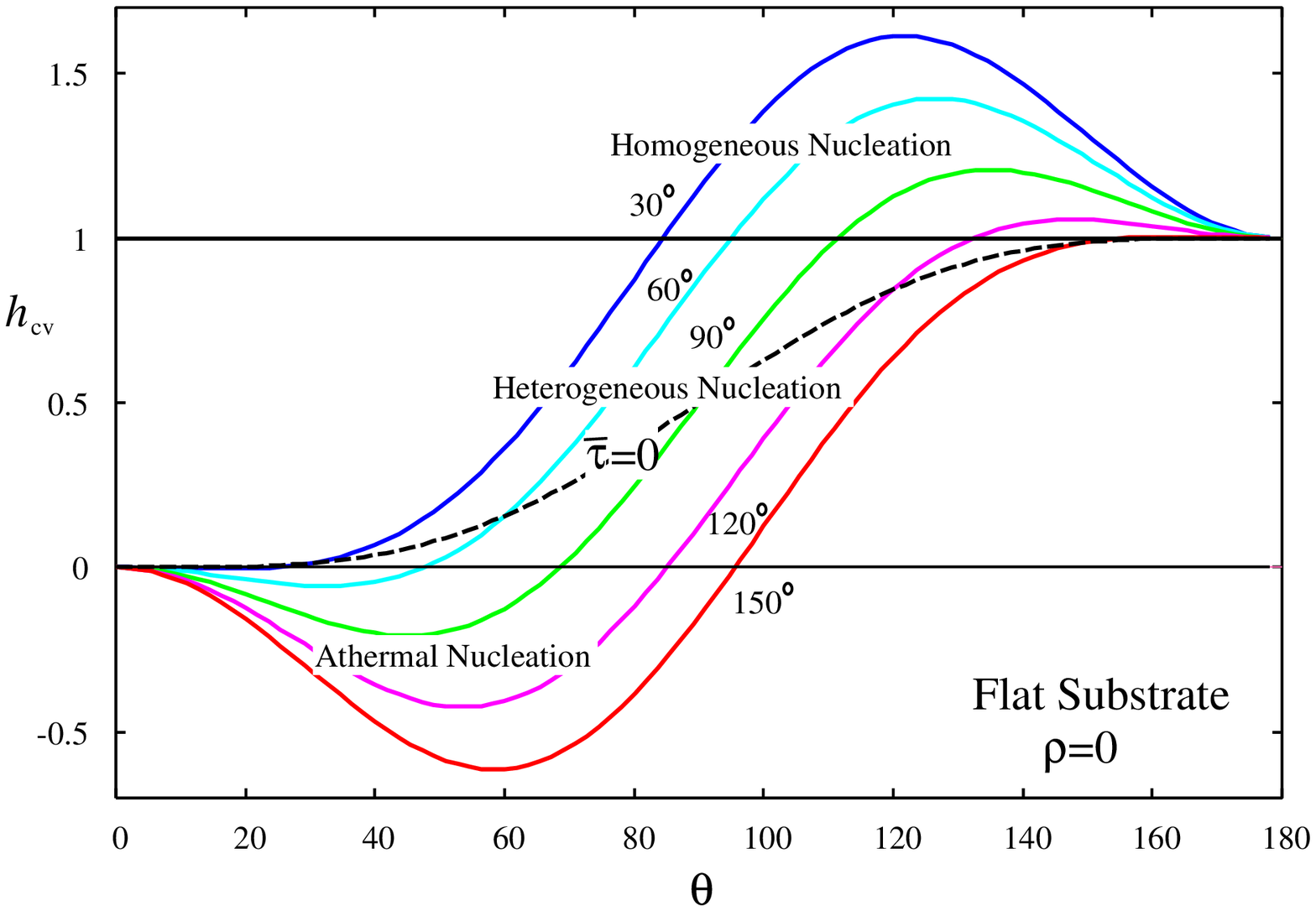}
\label{fig:LN6b}
}
\end{center}
\caption{
(a) Scaled line tension $\bar{\tau}_{\rm cv}\left(\rho,\theta\right)$ as a function of the contact angle $\theta$ for various Young's contact angle $\theta_{0} (=30^{\circ}, 60^{\circ}, 90^{\circ}, 120^{\circ}, 150^{\circ})$ for $\rho=0$, which corresponds to a flat substrate. The heterogeneous nucleation with a lens-shaped nucleus is favorable between the area sandwiched by the upper boundary $\bar{\tau}_{\rm u, cv}$ (upper black curve) and the lower boundary $\bar{\tau}_{\rm l, cv}$ (lower black curve).  (b) The corresponding scaled free energy $h_{\rm cv}\left(\rho,\theta\right)$.  The black broken curve indicates the free energy without line tension ($\tau=0$).  Therefore, the free-energy curves above this broken curve are the free energy when the line tension is positive ($\tau>0$).  The heterogeneous nucleation with a lens-shaped nucleus is favorable between the area sandwiched by two horizontal lines ($0\leq h_{\rm cv}\leq 1$). 
 } 
\label{fig:LN6}
\end{figure}

Figure~\ref{fig:LN6} shows the scaled line tension $\bar{\tau}_{\rm cv}\left(\rho=0,\theta\right)$ (Fig.~\ref{fig:LN6a}) and the corresponding scaled energy barrier $h_{\rm cv}\left(\rho=0,\theta\right)$ (Fig.~\ref{fig:LN6b}) of a nucleus on a flat substrate ($\rho=0$) as functions of the contact angle $\theta$ for different values of $\theta_{0}$.  We also show the upper bound $\bar{\tau}_{\rm u,cv}\left(\rho=0,\theta\right)$ and the lower bound $\bar{\tau}_{\rm l,cv}\left(\rho=0,\theta\right)$.  For a given line tension $\bar{\tau}$, the corresponding contact angle $\theta$ will be determined from the solution of equation
\begin{equation}
\bar{\tau}=\bar{\tau}_{\rm cv}\left(\rho,\theta\right).
\label{eq:LN36}
\end{equation}
Therefore, the intersection of a horizontal line $\bar{\tau}={\rm constant}$ and a curve $\bar{\tau}_{\rm cv}\left(\rho,\theta\right)$ will give the intrinsic contact angle $\theta$. 

When the contact angle $\theta$ increases, the contact-line length decreases as the radius $r_{*}$ is a constant given by the Young-Laplace equation (Eq.~(\ref{eq:LN12})).  Of course, when $\bar{\tau}=0$, the contact angle is given by the Young's contact angle $\theta=\theta_{0}$, as predicted from Eq.~(\ref{eq:LN35}).  Since the scaled line tension $\bar{\tau}$ approaches zero from Eq.~(\ref{eq:LN30}) as the size of the nucleus $r_{*}$ grows, the contact angle $\theta$ of the super-critical nucleus after crossing the energy barrier approaches the Young's contact angle $\theta_{0}$. When the line tension is positive and its magnitude $\bar{\tau}$ exceeds the upper bound $\bar{\tau}_{\rm u,cv}\left(\rho=0,\theta\right)$ (Fig.~\ref{fig:LN6a}), the scaled energy barrier exceeds the upper bound $h_{\rm cv}=1$ (Fig.~\ref{fig:LN6b}) for homogeneous nucleation.  Then, the heterogeneous nucleation becomes less probable and the homogeneous nucleation becomes dominant.  When the line tension is increased further, there will be no solution of Eq.~(\ref{eq:LN36}), which means that the lens-shaped heterogeneous nucleus is not the minimum of Helmholtz free energy anymore.  Then the heterogeneous nucleation will be inhibited and only the homogeneous nucleation will occur. 

Apparently, nucleation on a flat substrate is symmetric to hydrophilicity and hydrophobicity (Fig.~\ref{fig:LN6a}).  A nucleus with a positive line tension is more strongly attached to a hydrophilic substrate with smaller Young's contact angle $\theta_{0}$, such as $\theta_{0}=30^{\circ}$.  In this case, as the line tension increases, the contact angle begins to increase from the Young's contact angle $\theta_{0}=30^{\circ}$.  When the line tension reaches the upper bound $\bar{\tau}_{\rm u,cv}$, the homogeneous nucleation becomes favorable.

On the other hand, a nucleus with a negative line tension is favorably supported by the hydrophobic substrate with, for example, $\theta=150^{\circ}$ (Fig.~\ref{fig:LN6a}).  In this case, as the absolute magnitude of negative line tension increases, the contact angle $\theta$ decreases.  When the scaled line tension reaches the lower bound $\bar{\tau}_{\rm l,cv}$, the free energy becomes negative and the athermal nucleation will take over.   When the line tension is decreased further, there will be no solution of Eq.~(\ref{eq:LN36}), which means that the lens-shaped nucleus does not correspond to the minimum of Helmholtz free energy.  Then, the lens-shaped nucleus will not form and the nucleus will be a thin wetting layer to maximize the contact line length. The behavior of the contact angle is symmetrical for a neutral flat substrate with $\theta_{0}=90^{\circ}$ (Fig.~\ref{fig:LN6a}).  It can be hydrophilic ($\theta<90^{\circ}$) with a negative line tension or hydrophobic ($\theta>90^{\circ}$) with a positive line tension.

\begin{figure}[htbp]
\begin{center}
\subfigure[Scaled line tension $\bar{\tau}_{\rm cv}$ as a function of the contact angle $\theta$.]
{
\includegraphics[width=0.7\linewidth]{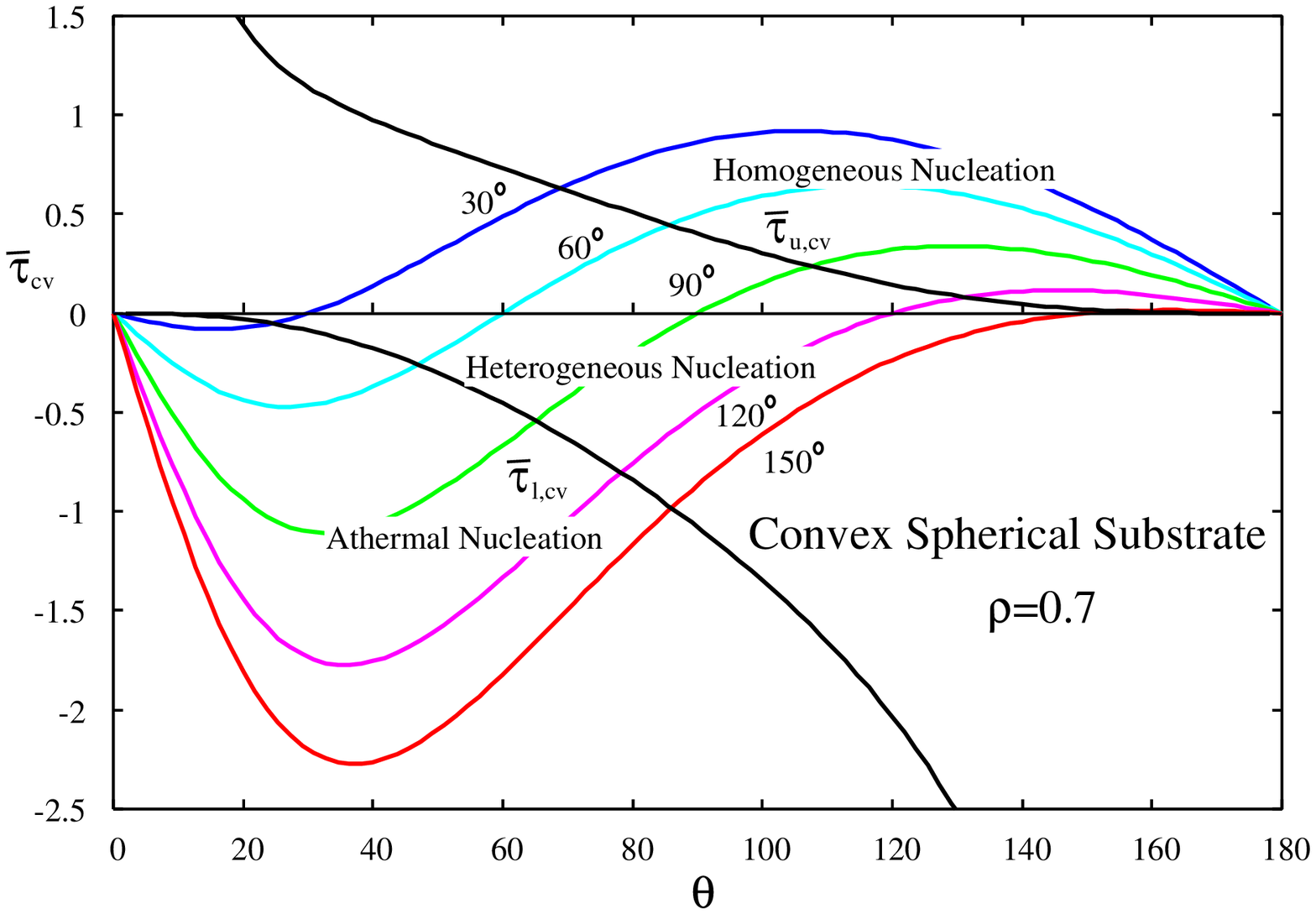}
\label{fig:LN7a}
}
\subfigure[Scaled energy barrier $h_{\rm cv}$ as a function of the contact angle $\theta$]
{
\includegraphics[width=0.7\linewidth]{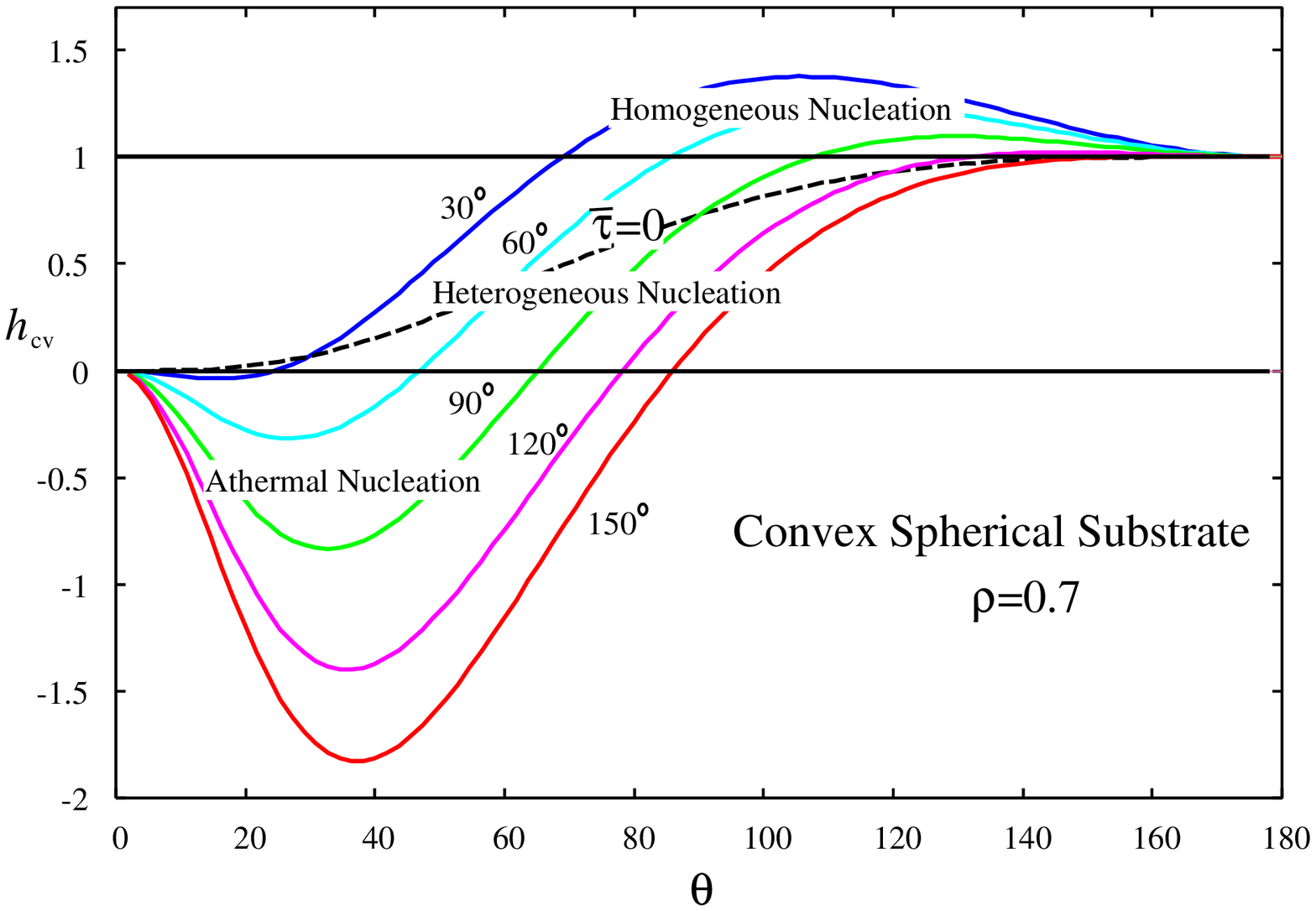}
\label{fig:LN7b}
}
\end{center}
\caption{
(a) Scale line tension $\bar{\tau}_{\rm cv}\left(\rho,\theta\right)$ of a nucleus on a convex substrate with $\rho=0.7$ as a function of the contact angle $\theta$ for various Young's contact angles $\theta_{0} (=30^{\circ}, 60^{\circ}, 90^{\circ}, 120^{\circ}, 150^{\circ})$.  All lines are shifted downward from Fig.~\ref{fig:LN6a}.  
(b) The corresponding scaled free energy $h_{\rm cv}\left(\rho,\theta\right)$.
 } 
\label{fig:LN7}
\end{figure}

As long as the radius $r_{*}$ of the critical nucleus is smaller than the radius $R$ of the spherical substrate, and $\rho<1$, the nucleus will remain on the upper hemisphere.  The situation does not differ much from that on a flat substrate.  Figure \ref{fig:LN7} shows the scaled line tension $\bar{\tau}_{\rm cv}$ and scaled energy barrier $h_{\rm cv}$ when $\rho=0.7$ for various values of $\theta_{0}$.  These figures are similar to those in Fig.~\ref{fig:LN6}. The contact angle $\theta$ increases as we increase the scaled line tension $\bar{\tau}$.  

There are several subtle differences between Fig.~\ref{fig:LN6} and Fig.~\ref{fig:LN7}.  First, the spherical substrate is less resistant to a positive line tension.  A lens-shaped nucleus will be less favorable on a spherical substrate than on a flat substrate.  In contrast, the spherical substrate is more resistant to a negative line tension.  If we increase the magnitude of negative line tension, the lens-shaped heterogeneous nucleus is more durable on the spherical substrate than on the flat substrate.  It can be intuitively imagined that the nucleus on a spherical substrate is less resistant to the tension that induces the shrinkage of the contact line and increase of the contact angle on the upper hemisphere.  Physically, this is because the contribution of the volume term $\Delta G_{\rm vol}^{*}$ on the convex substrate in Eq.~(\ref{eq:LN14}), which will lower the free energy, will be smaller than that on a flat substrate because the volume of the lens-shaped nucleus on a convex substrate is less than that of a spherical-cap nucleus on a flat substrate.  

\begin{figure}[htbp]
\begin{center}
\subfigure[Scaled line tension $\bar{\tau}_{\rm cv}$ as a function of the contact angle $\theta$.]
{
\includegraphics[width=0.7\linewidth]{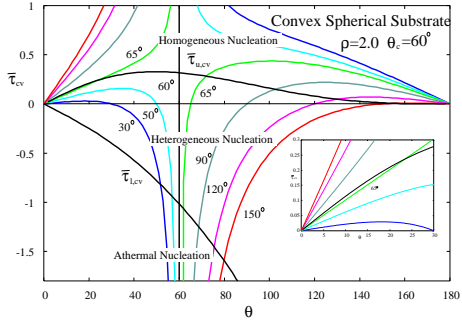}
\label{fig:LN8a}
}
\subfigure[Scaled energy barrier $h_{\rm cv}$ as a function of the contact angle $\theta$]
{
\includegraphics[width=0.7\linewidth]{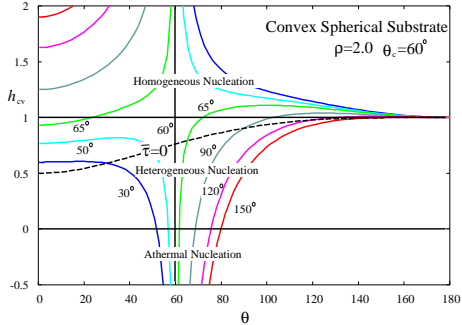}
\label{fig:LN8b}
}
\end{center}
\caption{
(a) The scaled line tension $\bar{\tau}_{\rm cv}\left(\rho,\theta\right)$  of a nucleus on a convex substrate with $\rho=2.0$ as a function of the contact angle $\theta$ for various Young's contact angles $\theta_{0} (=30^{\circ},  50^{\circ}, 60^{\circ}, 65^{\circ}, 90^{\circ}, 120^{\circ}, 150^{\circ})$. The critical angle is $\theta_{\rm c}=60^{\circ}$ on the basis of Eq.~(\ref{eq:LN41}).  Therefore, when the Young's contact angle is the critical angle ($\theta_{0}=\theta_{\rm c}=60^{\circ}$), the contact angle $\theta$ is pinned at $60^{\circ}$ (black solid vertical line).  The heterogeneous nucleation with a lens-shaped nucleus is favorable between the area sandwiched by the upper boundary $\bar{\tau}_{\rm u, cv}$ (upper black curve) and the lower boundary $\bar{\tau}_{\rm l, cv}$ (lower black curve).  (b) The corresponding scaled free energy $h_{\rm cv}\left(\rho,\theta\right)$. {The contact angle is pinned at $\theta=60^{\circ}$ when $\theta_{0}=60^{\circ}$ (black solid vertical line). }  Clearly, there exist two contact angles with the same energy barrier when $\theta<\theta_{\rm c}=60^{\circ}$ and when $\theta=65^{\circ}$ which is close to $\theta_{\rm c}$.  The heterogeneous nucleation with a lens-shaped nucleus is favorable between the area sandwiched by two horizontal lines ($0\leq h_{\rm cv}\leq 1$). 
 } 
\label{fig:LN8}
\end{figure}

\begin{figure}[htbp]
\begin{center}
\includegraphics[width=0.5\linewidth]{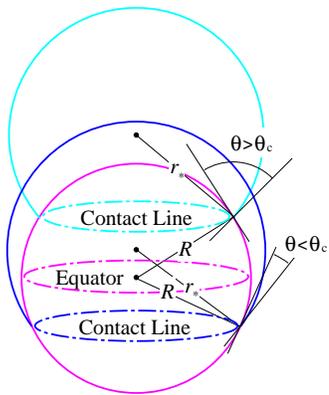}
\caption{
Two nuclei with the same radius $r_{*}>R$ ($\rho>1$).  The three-phase contact line is located on the upper hemisphere if $\theta>\theta_{\rm c}$ and on the lower hemisphere if $\theta<\theta_{\rm c}$.  On the upper hemisphere, the increase (decrease) in the contact-line length leads to an expansion (shrinkage) in the contact line, accompanied by a decrease (increase) in the contact angle.  On the lower hemisphere, however, the increase (decrease) of the contact-line results in the shrinkage (expansion) of the contact line accompanied by the increase (decrease) of the contact angle.  These two nuclei with the same radius $r_{*}$ and different contact angles $\theta$ may have the same free energy.
 }
\label{fig:LN9}
\end{center}
\end{figure}

The situation is more complicated once the critical radius $r_{*}$ of the nucleus exceeds the radius $R$ of the substrate ($\rho>1$). Figure \ref{fig:LN8} shows the scaled line tension $\bar{\tau}_{\rm cv}$ and the scaled energy barrier $h_{\rm cv}$ when $\rho=2.0$ for various values of $\theta_{0}$.  In this case, there exists a characteristic contact angle $\theta=\theta_{\rm c}$ given by
\begin{equation}
\theta_{\rm c}=\cos^{-1}\frac{1}{\rho},
\label{eq:LN37}
\end{equation}
which corresponds to Eq.~(\ref{eq:LN8}) where $\phi=90^{\circ}$ and the three-phase contact line coincides with the equator of the spherical substrate.  Therefore when $\theta<\theta_{\rm c}$, the three-phase contact line crosses the equator of the spherical substrate, moving from upper hemisphere to lower hemisphere (Fig.~\ref{fig:LN9}).   

If it happens that the Young's contact angle $\theta_{0}$, which characterizes the chemical nature of the substrate, coincides with the characteristic contact angle $\theta_{\rm c}$, which is determined solely from the size of substrate ($\theta_{0}=\theta_{\rm c}$), then the solution of the generalized Young equation (Eq.~(\ref{eq:LN3})) is given by $\theta=\theta_{0}$.  Therefore, when the Young's contact angle $\theta_{0}$ is the characteristic angle $\theta_{\rm c}$ given by Eq.~(\ref{eq:LN37}), the contact angle $\theta$ is pinned at the Young's contact angle $\theta_{0}=\theta_{\rm c}$, and the contact line stays on the equator, irrespective of the magnitude of scaled line tension $\bar{\tau}$.  In this case, as the line tension is increased, the contact angle is pinned at $\theta=\theta_{\rm c}=60^{\circ}$ (Fig.~\ref{fig:LN8a}, black vertical straight line).  When the line tension reaches the upper bound $\bar{\tau}_{\rm u,cv}$, the free energy of the lens-shaped heterogeneous nucleus exceeds that of the spherical homogeneous nucleus.  Then, the homogeneous nucleation rather than the heterogeneous nucleation becomes favorable.  At the contact angle $\theta_{0}=\theta_{\rm c}$, the line tension does not affect the contact angle $\theta$ (Fig.~\ref{fig:LN8a}), but it affects the energy barrier (Fig.~\ref{fig:LN8b}).  Since $\theta_{\rm c}<90^{\circ}$, the convex substrate must be hydrophilic so that the contact line crosses the equator. 

Note that $\theta\rightarrow 0^{\circ}$ limit of the free energy with $\tau=0$ (black broken curve) for $\rho>1$ in Fig.~\ref{fig:LN8b} is different from the same limit for $\rho<1$ in Fig.~\ref{fig:LN7b} since the critical nucleus with radius $r_{*}>R$ will completely wrap the spherical substrate of radius $R$.  In fact, from Eq.~(\ref{eq:LN16}) we have 
\begin{eqnarray}
f_{\rm cv}\left(\rho,\theta\rightarrow 0^{\circ}\right)
&=& 1-\frac{3}{\rho^{2}}+\frac{2}{\rho^{3}}\;\;\;\;\rho>1, \nonumber \\
&=& 0 \;\;\;\;\;\;\;\;\rho\leq 0.
\label{eq:LN38}
\end{eqnarray}
On the other hand
\begin{equation}
f_{\rm cv}\left(\rho,\theta\rightarrow 180^{\circ}\right)=1
\label{eq:LN39}
\end{equation}
always holds as shown in Fig.~\ref{fig:LN7b} and Fig.~\ref{fig:LN8b}. 

In contrast to Figs.~\ref{fig:LN6a} and \ref{fig:LN7a}, the lens-shaped nucleus can always exist when $\rho>1$ as Eq.~(\ref{eq:LN36}) can always have solutions for from Fig.~\ref{fig:LN8a} though the probability of occurrence might be very small.  Therefore, the homogeneous nucleation and the heterogeneous nucleation can coexist irrespective of the magnitude of the line tension when it is positive.  Similarly, the athermal nucleation with lens-shaped nucleus can coexist with an uniform wetting layer when the line tension is lower than the lower boundary  $\bar{\tau}_{\rm l, cv}$ 

The lens-shaped nucleus on a spherical substrate with $\rho>1$ (Fig.~\ref{fig:LN8a}) is less resistant to a positive line tension than that with $\rho<1$ (Fig.~\ref{fig:LN7a}).  The maximum line tension $\tau_{\rm u,cv}$ is smaller in Fig.~\ref{fig:LN8a} than in Fig.~\ref{fig:LN7a}; a lens-shaped heterogeneous nucleus is less favorable when $\rho>1$.   There always exist two branches for the same $\theta_{0}$ in Figs.~\ref{fig:LN8a} and \ref{fig:LN8b}.  The behavior of the contact angle $\theta$ with respect to the line tension $\bar{\tau}$ is different between the conditions $\theta_{0}>\theta_{\rm c}$ and $\theta_{0}<\theta_{\rm c}$.  When $\theta_{0}<\theta_{\rm c}$, the contact line is located on the lower hemisphere (Fig.~\ref{fig:LN9}). If the line tension is positive ($\bar{\tau}>0$), there will be two solutions of Eq.~(\ref{eq:LN36}) (see for example, Fig.~\ref{fig:LN8a}, where $\theta_{0}=30^{\circ}, 50^{\circ}$), one near the Young's contact angle $\theta_{0}$ and another near $0^{\circ}$.  The former contact angle increases and the latter contact angle decreases as the magnitude of positive line tension is increased.  There also exists a maximum, above which the Eq.~(\ref{eq:LN36}) has no solution when $\theta_{0}>\theta_{\rm c}$.  Then, only the homogeneous nucleation will be realized even though the line tension is smaller than the upper bound $\bar{\tau}_{\rm u, cv}$.   If the line tension is negative, the contact angle increases on increasing the magnitude of negative line tension (Fig.~\ref{fig:LN8a}). 

On the other hand, when $\theta_{0}>\theta_{\rm c}$ (Fig.~\ref{fig:LN8a}, $\theta_{0}=65^{\circ}-150^{\circ}$), the situation is similar to when $\rho=0$ in Fig.~\ref{fig:LN6a} and $\rho=0.7$ in Fig.~\ref{fig:LN7a}.  Now, the contact line is located on the upper hemisphere (Fig.~\ref{fig:LN9}), as in the case of $\rho<1$.  Increasing the line tension leads to a decrease in the nucleus perimeter.  Then, the contact angle $\theta$ increases as the line tension $\bar{\tau}$ is increased.  When it reaches the upper bound $\bar{\tau}_{\rm u,cv}$, the lens-shaped nucleus of heterogeneous nucleation becomes less favorable to a spherical nucleus of homogeneous nucleation with $\theta=180^{\circ}$ (Fig.~\ref{fig:LN8a}, $\theta_{0}=65^{\circ}-150^{\circ}$). 

When $\theta_{0}$ is very close to $\theta_{\rm c}$ ($\theta_{0}=65^{\circ}$), there exist two solutions (contact angles) of Eq.~(\ref{eq:LN36}) for the positive line tension $\bar{\tau}$ (Fig.~\ref{fig:LN8a},  $\theta_{0}=65^{\circ}$). The contact line of one solution is located on the upper hemisphere and another on the lower hemisphere (Fig.~\ref{fig:LN9}).  One solution, the contact line of which is located on the upper hemisphere, is $\theta=\theta_{0}$ at $\bar{\tau}=0$. The other solution, the contact line of which is located on the lower hemisphere, starts from $\theta=0^{\circ}$  at $\bar{\tau}=0$.  These two solutions have the same free energy (Fig.~\ref{fig:LN8b}). They both become less favorable to a spherical nucleus of homogeneous nucleation  when the line tension reaches the upper bound $\bar{\tau}_{\rm u,cv}$ (two green solid curve in Fig.~\ref{fig:LN8a} and in the inset). 

When $\rho>1$, the three-phase contact line can cross the equator of the spherical substrate.  The contact line is located on the upper hemisphere if $\theta_{0}>\theta_{\rm c}$, and on the lower hemisphere if $\theta_{0}<\theta_{\rm c}$.  The line tension cannot change the position of the contact line from the upper hemisphere to the lower hemisphere or from the lower hemisphere to the upper hemisphere because $\theta>\theta_{0}>\theta_{\rm c}$ and $\theta<\theta_{0}<\theta_{\rm c}$ always hold on the upper and lower hemisphere, respectively. Two nuclei with the same radius $r_{*}$ and different contact angles $\theta$, which are located on the upper and lower hemispheres (Fig.~\ref{fig:LN9}), may have the same free energy when the contact angle $\theta_{0}\sim \theta_{\rm c}$, which is true when $\theta_{0}=65^{\circ}$. Therefore, we will have two critical point of the nucleation, and we may expect parallel nucleation~\cite{Iwamatsu2012}.

\subsection{Nucleus on a concave spherical substrate}

The scaled free energy (energy barrier) of the heterogeneous critical nucleus nucleated on a concave spherical substrate of cavity is given by
\begin{equation}
h_{\rm cc}\left(\rho,\theta\right) = f_{\rm cc}\left(\rho,\theta\right)+\frac{3}{2}\bar{\tau}{\rm g}_{\rm cc}\left(\rho,\theta\right),
\label{eq:LN40}
\end{equation}
where $ f_{\rm cc}$ and ${\rm g}_{\rm cc}$ are defined in Eqs.~(\ref{eq:LN25}) and (\ref{eq:LN26}).

\begin{figure}[htbp]
\begin{center}
\subfigure[Energy barrier $h_{\rm cc}$ when the line tension is positive ($\bar{\tau}=0.1$)]
{
\includegraphics[width=0.7\linewidth]{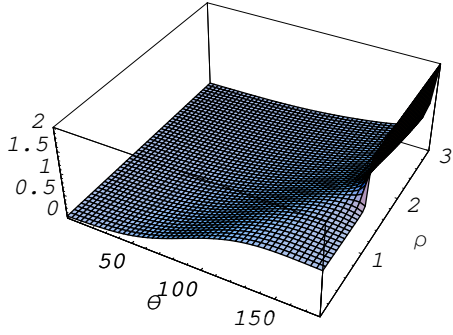}
\label{fig:LN10a}
}
\subfigure[Energy barrier $h_{\rm cc}$ when the line tension is negative ($\bar{\tau}=-0.1$)]
{
\includegraphics[width=0.7\linewidth]{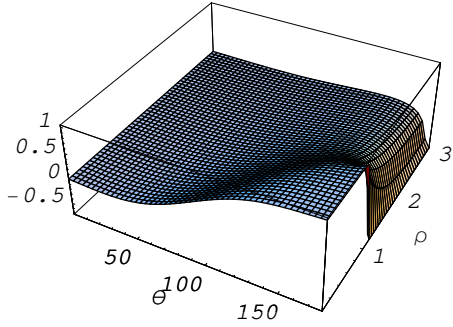}
\label{fig:LN10b}
}
\end{center}
\caption{
Scaled free energy (energy barrier) $h_{\rm cc}\left(\rho,\theta\right)$ of the critical nucleus on a concave spherical substrate as a function of the contact angle $\theta$ and size parameter $\rho$ when the scaled line tension is (a) positive ($\bar{\tau}=0.1$) and (b) negative ($\bar{\tau}=-0.1$). 
 } 
\label{fig:LN10}
\end{figure}

Figure \ref{fig:LN10} shows the scaled energy barrier  $h_{\rm cc}\left(\rho,\theta\right)$ as a functions of contact angle $\theta$ and size parameter $\rho$ for a heterogeneous nucleus in a spherical cavity when $\bar{\tau}=0.1>0$ (Fig.~\ref{fig:LN10a}) and $\bar{\tau}=-0.1<0$  (Fig.~\ref{fig:LN10b}).  It can be seen from Fig.~\ref{fig:LN10a} that the energy barrier exceeds $h_{\rm cc}=1$ when $\bar{\tau}>0$,  $\theta$ is large and $\rho$ is small.  Clearly, a small cavity of radius smaller than the radius of a critical nucleus cannot accommodate a spherical nucleus of homogeneous nucleation.  Therefore $\rho$ must be less than 1 when $h_{\rm cc}>1$ to realize homogeneous nucleation.  Otherwise, the inequality $h_{\rm cc}>1$ simply implies that a nucleus confined within a cavity has higher energy than the corresponding spherical nucleus of homogeneous nucleation.  On the other hand, when $\bar{\tau}<0$, $\theta$ is small and $\rho$ is large in Fig.~\ref{fig:LN10b}, and the scaled energy barrier becomes negative.  

The transition from heterogeneous-dominant nucleation to homogeneous-dominant nucleation occurs on a concave substrate when $h_{\rm cc}\left(\rho,\theta\right)=1$, which gives the upper bound of the scaled line tension
\begin{equation}
\bar{\tau}_{\rm u,cc}\left(\rho,\theta\right)=\frac{2\left(1- f_{\rm cc}\left(\rho,\theta\right)\right)}{3{\rm g}_{\rm cc}\left(\rho,\theta\right)},
\label{eq:LN41}
\end{equation}
which is derived from Eq.~(\ref{eq:LN33}) by replacing $f_{\rm cv}$ and ${\rm g}_{\rm cv}$ by  $f_{\rm cc}$ and ${\rm g}_{\rm cc}$.   The transition from activated nucleation to non-activated athermal nucleation occurs when  $h_{\rm cc}\left(\rho,\theta\right)=0$, which gives the lower bound
\begin{equation}
\bar{\tau}_{\rm l, cc}\left(\rho,\theta\right)=\frac{-2f_{\rm cc}\left(\rho,\theta\right)}{3{\rm g}_{\rm cc}\left(\rho,\theta\right)}.
\label{eq:LN42}
\end{equation}
The contact angle $\theta$ is determined from Eq.~(\ref{eq:LN23}).  Therefore, the scaled line tension $\bar{\tau}$ is given by
\begin{equation}
\bar{\tau}_{\rm cc}\left(\rho,\theta\right)=\frac{\left(\cos\theta_{0}-\cos\theta\right)\sin\theta}{1+\rho\cos\theta}.
\label{eq:LN43}
\end{equation}
where $\theta_{0}$ is defined by Eq.~(\ref{eq:LN4}).

Manipulation of the algebra shows that
\begin{eqnarray}
\bar{\tau}_{\rm cc}\left(\rho,\theta\right) &=& 
-\bar{\tau}_{\rm cv}\left(\rho,180^{\circ}-\theta\right), \nonumber \\
{\rm g}_{\rm cc}\left(\rho,\theta\right) &=& {\rm g}_{\rm cv}\left(\rho,180^{\circ}-\theta\right),
\label{eq:LN44} \\
f_{\rm cc}\left(\rho,\theta\right) &=& 1-f_{\rm cv}\left(\rho,180^{\circ}-\theta\right).
\nonumber
\end{eqnarray}
Then,
\begin{eqnarray}
\bar{\tau}_{\rm u, cc} &=& -\bar{\tau}_{\rm l, cv}, \nonumber \\
\bar{\tau}_{\rm l, cc} &=& -\bar{\tau}_{\rm u, cv}. 
\label{eq:LN45}
\end{eqnarray}
Therefore, the results for a concave substrate can be obtained by interchanging the role of hydrophilicity and hydrophobicity, and reversing the sign of the line tension ($\theta\rightarrow 180^{\circ}-\theta$ and $\tau \rightarrow -\tau$) in the results for a convex substrate.  The intrinsic contact angle $\theta$ is determined from
\begin{equation}
\bar{\tau}=\bar{\tau}_{\rm cc}\left(\rho,\theta\right)
\label{eq:LN46}
\end{equation}
similar to Eq.~(\ref{eq:LN36}).

\begin{figure}[htbp]
\begin{center}
\subfigure[Scaled line tension $\bar{\tau}_{\rm cc}$ as a function of the contact angle $\theta$.]
{
\includegraphics[width=0.7\linewidth]{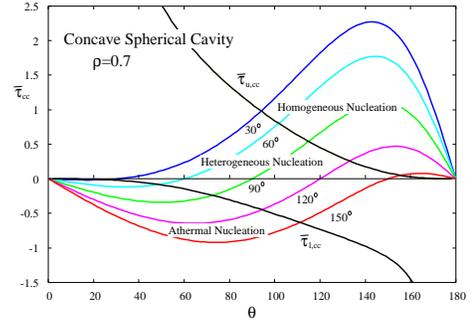}
\label{fig:LN11a}
}
\subfigure[Scaled energy barrier $h_{\rm cc}$ as a function of the contact angle $\theta$]
{
\includegraphics[width=0.7\linewidth]{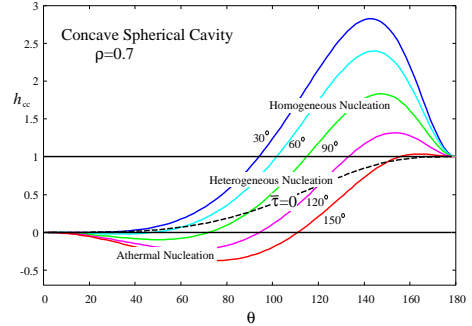}
\label{fig:LN11b}
}
\end{center}
\caption{
(a) Scaled line tension $\bar{\tau}_{\rm cc}\left(\rho,\theta\right)$ of the nucleus on a concave substrate with $\rho=0.7$ as a function of the contact angle $\theta$ for various Young's contact angles $\theta_{0} (=30^{\circ}, 60^{\circ}, 90^{\circ}, 120^{\circ}, 150^{\circ})$ .    (b) Corresponding scaled free energies $h_{\rm cc}\left(\rho,\theta\right)$.
}
\label{fig:LN11}
\end{figure}

As long as the critical radius $r_{*}$ of the nucleus is smaller than the radius $R$ of the spherical cavity, and $\rho<1$, the situation does not differ much from that for a flat or convex substrate with $\rho<1$ in Fig.~\ref{fig:LN7}.  Figure \ref{fig:LN11} shows the scaled line tension $\bar{\tau}_{\rm cc}$ and the scaled energy barrier $h_{\rm cc}$ when $\rho=0.7$ for various values of $\theta_{0}$.  These figures are similar to Figs.~\ref{fig:LN6} and \ref{fig:LN7}.  As noted in Eqs.~(\ref{eq:LN44}) and (\ref{eq:LN45}), Fig.~\ref{fig:LN11a} is obtained by rotating Fig.~\ref{fig:LN7a} clockwise by $180^{\circ}$.  In other words, Fig.~\ref{fig:LN11a} is obtained by changing the contact angle $\theta\rightarrow 180^{\circ}-\theta$ and $\bar{\tau}\rightarrow -\bar{\tau}$ in Fig.~\ref{fig:LN7a}.   Therefore, a convex substrate and concave substrate are symmetric in that a hydrophilic convex substrate with $\theta_{0}=30^{\circ}$ (Fig.~\ref{fig:LN7a}), for example, corresponds to a hydrophobic concave substrate with $\theta_{0}=180^{\circ}-30^{\circ}=150^{\circ}$ (Fig.~\ref{fig:LN11a}) and vice versa.  The positive line tension in the former case has the same effect as the negative line tension of the same magnitude in the latter case. 

There are several differences between Fig.~\ref{fig:LN11} and Fig.~\ref{fig:LN7}.  First, the concave substrate is more resistant to a positive line tension than to a negative line tension.  When the line tension is positive, the heterogeneous nucleation is favorable on a concave substrate (Fig.~\ref{fig:LN11a}) than on a convex  substrate (Fig.~\ref{fig:LN7a}) or on a flat substrate  (Fig.~\ref{fig:LN6a}).  In contrast, a concave substrate is less resistant to a negative line tension.  If we increase the absolute magnitude of negative line tension, the athermal nucleation rather than heterogeneous nucleation is easily realized, as the energy barrier becomes negative.  On the other hand, if we increase the positive line tension in Fig.~\ref{fig:LN11a}, the contact angle increases, as in Fig.~\ref{fig:LN7a}.  As soon as the line tension reaches the upper bound $\bar{\tau}_{\rm u, cc}$, the homogeneous nucleation rather than the heterogeneous nucleation becomes favorable. 

\begin{figure}[htbp]
\begin{center}
\subfigure[Scaled line tension $\bar{\tau}_{\rm cc}$ as a function of the contact angle $\theta$.]
{
\includegraphics[width=0.7\linewidth]{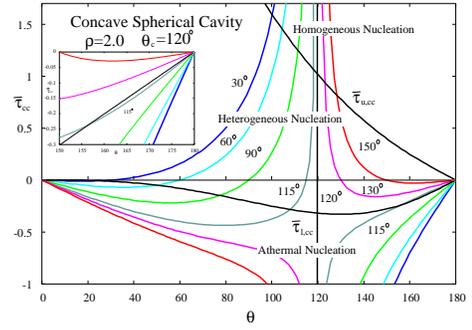}
\label{fig:LN12a}
}
\subfigure[Scaled energy barrier $h_{\rm cc}$ as a function of the contact angle $\theta$]
{
\includegraphics[width=0.7\linewidth]{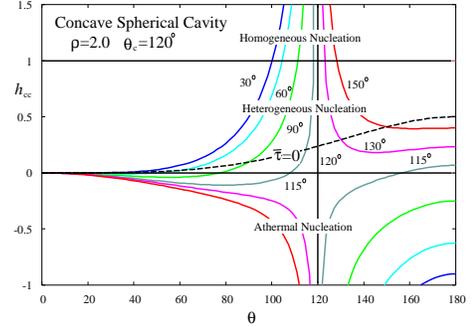}
\label{fig:LN12b}
}
\end{center}
\caption{
(a) Scaled line tension $\bar{\tau}_{\rm cc}\left(\rho,\theta\right)$ as a function of the contact angle $\theta$ for various Young's contact angle $\theta_{0} (=30^{\circ},  60^{\circ}, 90^{\circ}, 115^{\circ},  120^{\circ}, 130^{\circ}, 150^{\circ})$ when $\rho=2.0$. The characteristic angle is $\theta_{\rm c}=120^{\circ}$ on the basis of Eq.~(\ref{eq:LN44}).  Therefore, when the Young's contact angle is $\theta_{0}=\theta_{\rm c}=120^{\circ}$, the contact angle $\theta$ is pinned at $120^{\circ}$. In this case, the upper bound $\bar{\tau}_{\rm u, cc}$ is meaningless as the cavity is too small to accommodate the entire volume of the nucleus.  (b) The corresponding scaled free energy $h_{\rm cc}\left(\rho,\theta\right)$. 
 } 
\label{fig:LN12}
\end{figure}

The situation is more complicated once the critical radius $r_{*}$ of the nucleus exceeds the radius $R$ of the concave substrate ($\rho>1$), as in the case of the convex substrate.  Furthermore, there is large difference between the concave and convex substrates because the concave cavity with radius $R$ cannot accommodate the spherical homogeneous nucleus with radius $r_{*}>R$. Therefore, the homogeneous nucleation with the contact angle $180^{\circ}$ is inhibited.  The upper bound in Eq.~(\ref{eq:LN41}) has no meaning but the lower bound in Eq.~(\ref{eq:LN42}) still exists when $\rho>1$.

This fact is reflected on the $\theta\rightarrow 180^{\circ}$ limit of the free energy with $\tau=0$ (black broken curve) for $\rho>1$ in Fig.~\ref{fig:LN12b} which is different from the same limit for $\rho<1$ in Fig.~\ref{fig:LN11b}.  In fact, from Eq.~(\ref{eq:LN25}) we have 
\begin{eqnarray}
f_{\rm cc}\left(\rho,\theta\rightarrow 180^{\circ}\right)
&=& \frac{3}{\rho^{2}}-\frac{2}{\rho^{3}}\;\;\;\;\rho>1, \nonumber \\
&=& 1 \;\;\;\;\;\;\;\;\rho\leq 1.
\label{eq:LN47}
\end{eqnarray}
On the other hand
\begin{equation}
f_{\rm cc}\left(\rho,\theta\rightarrow 0^{\circ}\right)=0
\label{eq:LN48}
\end{equation}
always holds as shown in Fig.~\ref{fig:LN11b} and Fig.~\ref{fig:LN12b}.

With the exception of the homogeneous nucleation, the situation is  similar to the case of the convex substrate when $\rho>1$.  Figure \ref{fig:LN12} shows the scaled line tension $\bar{\tau}_{\rm cc}$ and the scaled energy barrier $h_{\rm cc}$ for various values of $\theta_{0}$ and $\rho=2.0$.   Again, Fig.~\ref{fig:LN12a} is obtained by rotating Fig.~\ref{fig:LN8a} clockwise by $180^{\circ}$.  The hydrophobicity and the hydrophilicity interchange their roles in concave and convex substrates.  In this case, there also exists a characteristic contact angle $\theta=\theta_{\rm c}$ for the droplet on a concave substrate given by
\begin{equation}
\theta_{\rm c}=\cos^{-1}\frac{-1}{\rho},
\label{eq:LN49}
\end{equation}
which corresponds to Eq.~(\ref{eq:LN24}).  Therefore, $\theta_{\rm c}>90^{\circ}$, and the substrate must be hydrophobic so that contact line crosses the equator of the substrate at $\phi=90^{\circ}$.

One solution of Eq.~(\ref{eq:LN23}) is $\theta=\theta_{0}=\theta_{\rm c}$.  Therefore, when the Young's contact angle is given by the characteristic contact angle ($\theta_{0}=\theta_{\rm c}$), the contact angle $\theta$ is pinned at $\theta_{\rm c}$ irrespective of the magnitude of scaled line tension $\bar{\tau}$.  As the line tension is changed, the contact angle is pinned at $\theta=\theta_{0}=120^{\circ}$ (Fig.~\ref{fig:LN12a}, black straight line) as $\theta_{\rm c}=120^{\circ}$ when $\rho=2.0$ (from Eq.~(\ref{eq:LN49})).    

\begin{figure}[htbp]
\begin{center}
\includegraphics[width=0.5\linewidth]{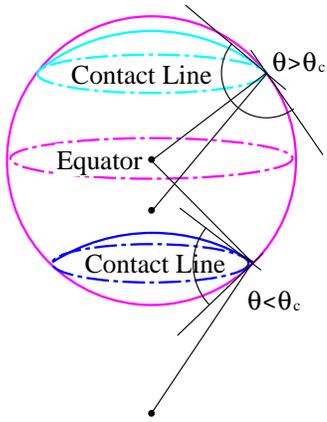}
\caption{
Two nuclei with the same radius $r_{*}>R$ ($\rho>1$).  The three-phase contact line is located on the upper hemisphere if $\theta>\theta_{\rm c}$, and on the lower hemisphere if $\theta<\theta_{\rm c}$. These two nuclei with the same radius $r_{*}$ and different contact angles $\theta$ may have the same free energy.
 }
\label{fig:LN13}
\end{center}
\end{figure}

The behavior of the contact angle $\theta$ in Fig.~\ref{fig:LN12a} as a function of the line tension $\bar{\tau}_{\rm cc}$ is different between the condition $\theta_{0}>\theta_{\rm c}$ and $\theta_{0}<\theta_{\rm c}$, as in the case of a convex substrate. When $\theta_{0}<\theta_{\rm c}$, the situation is similar to that when $\rho=0.7$ in Fig.~\ref{fig:LN11a} and when $\theta_{0}>\theta_{\rm c}$ in Fig.~\ref{fig:LN8a}.  In this case, the contact line is located on the lower hemisphere (Fig.~\ref{fig:LN13}). The contact angle $\theta$ increases as the line tension $\bar{\tau}$ is increased.  When $\theta_{0}>\theta_{\rm c}$, on the other hand, the contact line is located on the upper hemisphere  (Fig.~\ref{fig:LN13}), and the contact angle $\theta$ decreases as the line tension $\bar{\tau}$ is increased. Again, the line tension cannot change the position of the contact line from the upper hemisphere to the lower hemisphere or from the lower hemisphere to the upper hemisphere, because $\theta_{0}>\theta>\theta_{c}$ and $\theta_{\rm c}>\theta>\theta_{0}$ always hold on the upper and lower hemispheres, respectively, of the concave substrate. 

When $\theta_{0}$ is very close to $\theta_{\rm c}$ (Fig.~\ref{fig:LN12}, grey solid curve for $\theta_{0}=115^{\circ}$), there exist two solutions of Eq.~(\ref{eq:LN46}) for the same negative line tension $\bar{\tau}$.  One solution locates on the lower hemisphere ($\theta<\theta_{\rm c}$) and the other solution locates on the upper hemisphere ($\theta>\theta_{\rm c}$).  These two critical nuclei with different contact angles have the same free energy (Figs.~\ref{fig:LN12b} and \ref{fig:LN13}).  There also exist two solutions of Eq.~(\ref{eq:LN46}) on the same upper hemisphere ($\theta>\theta_{\rm c}$) when the line tension is negative ($\bar{\tau}<0$) and $\theta_{0}>\theta_{\rm c}$ (Fig.~\ref{fig:LN12}, $\theta_{\rm c}=130^{\circ}, 150^{\circ}$).  Clearly, these two critical nuclei with different contact angles also have the same free energy (Figs.~\ref{fig:LN12b}).  Therefore, we may expect parallel nucleation~\cite{Iwamatsu2012} again.

\section{\label{sec:sec5} Conclusion}

In this study, we considered the line-tension-induced scenario of heterogeneous nucleation by considering the free energy of a lens-shaped critical nucleus nucleated on a spherical substrate and on the wall of a spherical cavity.  The generalized Young equation~\cite{Guzzardi2007,Hienola2007} was rederived by minimizing the Helmholtz free energy to determine the contact angle.  Then, we determined the critical radius and the work of formation (nucleation barrier) of the critical nucleus by using the generalized Young equation and by maximizing the Gibbs free energy.  The work of formation consists of the usual volume contribution and a line contribution due to the line tension, which was examined in our previous paper~\cite{Iwamatsu2015}.  Using the generalized Young equation, we studied the contact angle of a lens-shaped nucleus as a function of line tension.  As long as the contact line remains on the upper hemisphere of a spherical substrate or the lower hemisphere of a spherical cavity, the line-tension dependence of the contact angle on a convex or concave substrate is similar to that on a flat substrate.  Once the contact line can cross the equator of a sphere or cavity, a more complex behavior is observed.  Now, the line-tension dependence of the contact angle is different on the upper hemisphere from that on the lower hemisphere.  The contact line cannot cross the equator even if we change the magnitude of line tension.  

The scenario of heterogeneous nucleation of a lens-shaped nucleus was examined by comparing the Gibbs free energy of a lens-shaped nucleus with that of a spherical nucleus of the same critical radius of homogeneous nucleation.  On increasing the {\it positive line tension}, the free energy can exceed that of homogeneous nucleation.  Then, the homogeneous nucleation is more preferable to the heterogeneous nucleation.  On the other hand, on increasing the magnitude of {\it negative line tension}, the free energy barrier becomes negative.  Then, the nucleation becomes athermal nucleation~\cite{Quested2005} without thermal activation. 

Although, we have considered only a critical nucleus that is intrinsically metastable, it is possible to study a nonvolatile nucleus on a convex or concave substrate by following the work of Widom~\cite{Widom1995}.  However, the meaning of the scaled line tension becomes obscure for such a nucleus, although it is clearly defined in this work as the radius of nucleus is fixed by the Young-Laplace equation. 

We used a spherical lens-shaped nucleus model and assumed that the liquid-solid interaction is given by short-ranged contact interaction represented by the surface tension.  However, it is possible to include long-ranged liquid-solid interaction by using the concept of disjoining pressure.  Several studies~\cite{Hage1984,Kuni1996,Tsekov2000,Iwamatsu2013} have already examined the effect of the long-range force.  However, most of those studies have focused on the case of complete wetting with the contact angle $\theta=0^{\circ}$ for a spherical nucleus surrounding a spherical substrate~\cite{Kuni1996,Bieker1998,Tsekov2000,Bykov2006}, rather than a lens-shaped nucleus with a three-phase contact line.  Although there are numerous studies on lens-shaped nuclei on flat substrates in the presence of long-ranged disjoining pressure~\cite{Yeh1999,Zhang2002a,Starov2004,Iwamatsu2011}, investigations on lens-shaped nuclei on spherical substrates are scarce, as they have to rely on numerical studies~\cite{Dobbs1992}.  In addition, we have assumed that the liquid-vapor interface is sharp.  There are several studies~\cite{Padilla2001,Ghosh2013} that include diffuse interfaces. The results in this paper can serve as a guide in the development of more realistic models of lens-shaped nuclei on curved surfaces that includes the disjoining pressure and diffuse-interface effect.  In order to include diffuse-interface effect, however, the line-tension may cause problems~\cite{Schimmele2007}, which will be more troublesome on a curved substrate.


\end{document}